\documentclass[seceq]{ptptex}

\newcommand{\beq}{\begin{equation}}
\newcommand{\beqa}{\begin{eqnarray}}
\newcommand{\eeq}{\end{equation}}
\newcommand{\eeqa}{\end{eqnarray}}

\newcommand{\siml}{\lesssim}

\title{
Coherent Radiation in Gamma-Ray Bursts and Relativistic Collisionless Shocks%
}

\author{
Kunihito \textsc{Ioka}%
}

\inst{
Department of Physics, Kyoto University, Kyoto 606-8502, Japan
}

\abst{
We suggest that coherent radiation may occur in relativistic
collisionless shocks via two-stream Weibel instabilities. The
coherence amplifies the radiation power by many orders [$\sim 10^{12}$
in Gamma-Ray Bursts (GRBs)] and particles cool very fast before being
randomized. We imply (1) GRBs accompany strong infrared emission, (2)
protons efficiently transfer energy to electrons and (3) prompt GRBs
might be the upscattered coherent radiation.
}

\begin{document}

\maketitle

\section{Introduction} 

Gamma-Ray Bursts (GRBs) are believed to be produced by the dissipation
of the kinetic energy of a relativistic outflow, energized by a central
compact object, via collisionless shocks.\cite{rees05} Although the
leading GRB mechanism is synchrotron emission from accelerated
electrons, many problems remain unsolved and the true mechanism is
far from understood.\cite{zhang04} In view of recent
observational discoveries, such as the X-ray
flashes,\cite{heise01,yamazaki02,ioka01} the prompt infrared-optical
emission,\cite{blake05,vestrand05} and the empirical relations of the
peak energy of GRBs that may enable precision
cosmology,\cite{amati02,yonetoku04,yamazaki04} it is high time to reveal
the GRB emission mechanism, hopefully from the first principle. One of
the most promising approaches is to investigate the kinematics of the
relativistic collisionless shocks.\cite{medvedev99,silva03,kato05}

In relativistic collisionless shocks, plasma particles are initially
interpenetrating. The velocity distribution is anisotropic and the
anisotropy drives the Weibel two-stream instability
\cite{weibel59,fried59,kazimura98,medvedev99}. The intersecting particles are
deflected by a magnetic field perturbation so that the currents making
the magnetic fields increase. As a result the magnetic fields
perpendicular to the shock propagation are amplified and many current
filaments (cylindrical beams) parallel to the shock propagation are
generated.\cite{lee73} Each beam carries a current producing a magnetic
field around itself. Since like currents attract each other, they merge
and grow in size \cite{silva03,frederiksen04,medvedev05}. The magnetic
fields also grow and reach the maximum when the current growth saturates
at the Alfv$\acute{\rm e}$n limiting current \cite{kato05} and the
particles are randomized.

In this Letter we will add one more key to the above picture by
suggesting that current filaments could emit coherent radiation
before particles are randomized. A filament emits radiation because
charged particles in the filament have a common acceleration when the
filament is curved in order to merge. Many particles accelerate in the
same way, so that radiation from $N$ particles are added coherently
(i.e., $E \sim N E_{1}$) and the radiation power ($\propto |E|^2$)
becomes $\propto N^2$ rather than $\propto N$.\footnote{Our mechanism of
coherent radiation is different from that of Ref.~\citen{sagiv02}.} The
radiation power is amplified by many orders ($\sim 10^{12}$ in GRBs) and
particles may lose almost all energy within the turnaround radius. This
coherent radiation may have significant impacts on the GRB model: (1)
the GRBs may be associated with strong infrared emission, (2) the proton
energy may be efficiently transfered to electrons, and (3) even the GRB
itself might be the upscattered coherent radiation.

\section{Coherent radiation in relativistic collisionless shocks}

To model a single current filament, we consider particles with charge
$q$ and mass $m$ running in a filament with a velocity $c \beta = c
(1-\gamma^{-2})^{1/2} \sim c[1-(1/2 \gamma^{2})]$. The filament has a
radius $\lambda$ and is bent with a curvature radius $r>\lambda$.
Although this is a simple model, we can quantify the plausibility of the
coherent radiation. All quantities are measured in the shocked frame.

According to the linear theory of the Weibel
instability\cite{medvedev99,milos05}, the filament radius is about the
order of the plasma skin depth\footnote{More precisely the filament
radius depends also on the velocity distribution perpendicular to the
propagation.\cite{medvedev99,milos05}
The radius also increases when current filaments merge.}
\beqa
\lambda \sim \left(\frac{\pi m c^2}{q^2 n}\right)^{1/2}
\sim 3.3 n_{12}^{-1/2} (m/m_e)^{1/2} \ {\rm cm},
\label{eq:psd}
\eeqa
where $m_e$ is the electron mass, and we adopt the particle number
density $n=10^{12}n_{12}$ cm$^{-3}$ as typical parameters for the
internal shocks of the GRBs. (Note that the possibility of coherent
radiation does not depend on the density $n$. See below.)

The typical curvature radius $r$ of the filament would be about or larger than
the Larmor radius
\beqa
r_L = \frac{\gamma m c^2}{q B}
\sim 12  \epsilon_{B,-2}^{-1/2} n_{12}^{-1/2} \gamma_1^{1/2}
(m/m_e)^{1/2} \ {\rm cm},
\label{eq:rl}
\eeqa
where $\gamma=10 \gamma_1$ is a typical relative Lorentz factor
in the GRB internal shocks\footnote{Before the internal shocks each
ejecta would be cold because of the adiabatic expansion.} and
$\epsilon_B= B^2/8\pi n \gamma m c^2=10^{-2} \epsilon_{B,-2}$ is a
fraction of the magnetic energy. The curvature radius $r$ could be
larger than $r_L$, for example, in the case of a proton filament
Debye-shielded by electrons. [See Fig.~1 of Ref.~\citen{hededal04} and
Fig.~2 of Ref.~\citen{frederiksen04}.] So we leave the curvature radius
$r$ as a free parameter and consider two cases:
\beqa
(A)\ r > \gamma^2 \lambda, \quad (B)\ \gamma^2 \lambda \ge r> \lambda.
\eeqa
Note that $r_L \sim \gamma^{1/2} \lambda$ and hence 
$r \ge r_L >\lambda$ for relativistic shocks 
$\gamma \gg 1$ regardless of the density $n$ and the particle mass $m$
from equations (\ref{eq:psd}) and (\ref{eq:rl}).

Given the curvature radius $r$ of the filament, we can estimate the
radiation power from a single particle $P = 2 c q^2 \beta^4 \gamma^4/3
r^2$ and thereby the cooling length (without a coherent effect)
\beqa
l_1 \sim \frac{\gamma m c^3}{P} \sim \frac{3 r^2 m c^2}{2 q^2 \beta^4
\gamma^3}
\sim 5.3 \times 10^{13} r_2^2 \gamma_1^{-3} (m/m_e)
\ {\rm cm},
\label{eq:l1}
\eeqa
where $r=10^2 r_2$ cm. The cooling length $l_1$ is much larger than the
other scales, $r$ and $\lambda$, for typical parameters. However, if $N$
particles radiate coherently, the cooling length could be $N$ times
shorter ($l_N \sim l_1/N$) because the mass $m$ and charge $q$ in
equation (\ref{eq:l1}) $l_1 \propto m/q^2$ are replaced by $Nm$ and
$Nq$.
Coherent radiation has been also discussed in the context of
pulsars\cite{goldreich71} and particle
accelerators\cite{michel82,nakazato89}.

\subsection{The case (A) $r>\gamma^2 \lambda$}

In order to estimate the possible number of particles that can radiate
coherently $N$, we remember that a relativistic particle emits radiation
ahead into a cone of an angular size\cite{rybicki79} 
\beqa
\Delta \theta \sim \gamma^{-1}.
\eeqa
Because of this relativistic beaming, an observer will see radiation
from the particle's path length of $\Delta s \sim r \Delta \theta \sim
r/\gamma$. After passing $\Delta s$ the radiation extends to $\sim r
(\Delta \theta)^2 \sim r/\gamma^2$ in the perpendicular direction to the
propagation. Since this ($\sim r/\gamma^2$) is larger than the filament
radius $\lambda < r/\gamma^2$ in the case (A), radiation from the inside
of the filament may be coherently superimposed. In the propagation
direction, the front and back of the radiation is separated by $\sim r
(\Delta \theta)^3 \sim r/\gamma^3$ after passing $\Delta s \sim r \Delta
\theta$. Then the characteristic frequency of the radiation is
\beqa
\nu_* \sim \frac{c}{r(\Delta \theta)^3} \sim \frac{c \gamma^3}{r}
\sim 3.0 \times 10^{11} r_2^{-1} \gamma_1^{3}
\ {\rm Hz},
\label{eq:nu*}
\eeqa
and particles within the wave length $\sim r (\Delta \theta)^3 \sim
r/\gamma^3$ may be coherent. Therefore the possible number of coherent
particles is given by
\beqa
N \sim n (\pi \lambda^2) [r (\Delta \theta)^3]
\sim 3.5 \times 10^{12} r_2 \gamma_{1}^{-3} (m/m_e),
\label{eq:N}
\eeqa
and the cooling length may be as short as
\beqa
\frac{l_N}{r} \sim \frac{l_1/N}{r}
\sim \frac{3}{2 \pi^2 \beta^4} \sim 0.15.
\label{eq:lN}
\eeqa
Surprisingly particles may emit almost all energy within a very short
distance $\sim 0.1 r \sim 10 r_2$ cm before turning $\pi$ radian! An
interesting point is that the ratio $l_N/r$ is just a constant and does
not depend on other parameters if $\gamma \gg 1$.

\subsection{The case (B) $\gamma^2 \lambda \ge r > \lambda$}

If the frequency of interest is the characteristic frequency $\nu_* \sim
c \gamma^3/r$, not all particles within the filament radius can be
coherent since the perpendicular extent of the radiation $\sim r (\Delta
\theta)^2 \sim r/\gamma^2$ is smaller than the filament radius $\lambda
> r/\gamma^2$. However, if the frequency of interest $\nu$ is less than
$\nu_*$, the radiation is beamed into a wider angular size
\beqa
\Delta \theta \sim \gamma^{-1} (\nu/\nu_*)^{-1/3},
\label{eq:thB}
\eeqa
than $\sim \gamma^{-1}$. Then, at a frequency lower than
\beqa
\nu \sim \nu_* \left(\frac{r}{\gamma^2 \lambda}\right)^{3/2},
\label{eq:nu}
\eeqa
the radiation from the inside of the filament may be coherent since $r
(\Delta \theta)^2 > \lambda$. The wave length of the radiation is
$\sim r(\Delta \theta)^3 \sim r \gamma^{-3} (\nu/\nu_*)^{-1}$. 
[Note that this is less than the plasma skin depth $\lambda$
at a frequency in equation (\ref{eq:nu}).] 
Therefore
the possible number of coherent particles $N \sim n (\pi \lambda^2) [r
(\Delta \theta)^3]$ is larger than that in equation (\ref{eq:N}) by
$\sim (\nu/\nu_*)^{-1}$. On the other hand the incoherent cooling length
$l_1$ should be multiplied by $\sim (\nu/\nu_*)^{-4/3}$ because the
radiation power from a single particle at a frequency $\nu$ is $\sim
(\nu/\nu_*)^{4/3}$ times smaller than that at $\nu_*$. Consequently,
replacing $l_1$ and $N$ by $l_1 (\nu/\nu_*)^{-4/3}$ and $N
(\nu/\nu_*)^{-1}$ in equation (\ref{eq:lN}), we obtain 
\beqa
\frac{l_N}{r} \sim 0.15 \left(\frac{\nu}{\nu_*}\right)^{-1/3} 
\sim 0.15 \left(\frac{\gamma^2 \lambda}{r}\right)^{1/2},
\label{eq:lNB}
\eeqa
where we use equation (\ref{eq:nu}) in the last equality. Since the
curvature radius is larger than the Larmor radius $r > r_L \sim
\gamma^{1/2} \lambda$, we have $l_N/r < 0.1 \gamma^{3/4}$. Therefore
particles can lose their energy within $r$ as long as $\gamma \siml 20$
in the case (B).
Note that coherent radiation in the case (B) is less efficient than
that in the case (A).

\subsection{Radiative instability}

In order for the radiation to be coherent, the filament needs to be
clumpy with the scale of the wave length $\sim r (\Delta \theta)^3$.
Otherwise the particle distribution in the filament is random and
the emission is added incoherently. Since the Weibel instability only
makes structures perpendicular to the propagation, other mechanisms are
necessary to bunch the filament.

A clumpy filament may be produced
spontaneously by the radiative instability.\cite{goldreich71} A density
perturbation generates some coherent radiation and the interaction of
this radiation with particles may amplify the density perturbation.

Goldreich and Keeley\cite{goldreich71} studied the stability of a ring
of monoenergetic relativistic particles and applied to pulsars. They
found the unstable condition as
\beqa
\left[\gamma^3 (\Delta \theta)^3\right]^{4/3} 
\left(\frac{\gamma^3}{\pi \lambda^2 r n}\right)
\left(\frac{\gamma^2 q^2}{r m c^2}\right)
<1,
\eeqa
which gives $8 \times 10^{-26} \gamma_1^{5} r_2^{-2} (\nu/\nu_*)^{-4/3}
(m/m_e)^{-2}<1$ and is well satisfied in a large parameter space. They
also derived the growth length $d$ as
\beqa
\frac{d}{r} \sim \left(\frac{r m c^2}{q^2}
\frac{\gamma^3}{\pi \lambda^2 r n}\right)^{1/2}
(\Delta \theta)^{2}
\sim \frac{1}{\pi \gamma^{1/2}}
\left(\frac{\nu}{\nu_*}\right)^{-2/3}.
\eeqa
In the case (A) we have $\nu \sim \nu_*$ and so $d/r \sim 1/\pi
\gamma^{1/2} < 1$. Therefore the density fluctuation could be amplified
within the curvature radius $r$ for the coherent radiation to be
efficient. In the case (B) we have $d/r \sim \gamma^{3/2} \lambda/\pi r
< \gamma^{1/2}/\pi$ where we use equation (\ref{eq:nu}) and $r > r_L
\sim \gamma^{1/2} \lambda$. So the density fluctuation may grow
depending on the values of $r$ and $\gamma$.

\section{Possible GRB scenarios}

In the previous section we suggest a possibility that particles could
lose almost all energy in the relativistic collisionless shocks very
fast by emitting coherent radiation. Let us consider implications of
this possibility for the GRB models.

The first possibility is that strong infrared emission may be associated
with GRBs. The Weibel instability grows faster for electrons than for
protons, and electrons may emit almost all energy by coherent
radiation. Then the radiation may carry $m_e/m_p \sim 10^{-3}$ of all
kinetic energy, and has a frequency about
\beqa
\nu_{*,{\rm obs}} \sim \Gamma \nu_*
\sim 3 \times 10^{13} r_2^{-1} \gamma_1^{3} \Gamma_2
\ {\rm Hz},
\eeqa
where $\nu_*$ is given by equation (\ref{eq:nu*}) and $\Gamma=10^2
\Gamma_2$ is the Lorentz factor of the shocked ejecta. The coherent
radiation may stand out above the prompt GRB, which has only $\sim
\nu_{*,{\rm obs}}/100 {\rm keV} \sim 10^{-6}$ of all kinetic energy at
the infrared $\nu_{*,{\rm obs}}$.\footnote{Note that the synchrotron
self-absorption seems to be absent at least in the prompt
infrared-optical emission of GRB
041219a.\cite{blake05,vestrand05,fan05}}

Second the coherent radiation could efficiently transfer energy from
protons to electrons. Before being shocked protons carry $m_p/m_e\sim
10^3$ times more energy than electrons. In the standard GRB
model\cite{rees05} we assume that a large fraction of the proton energy
is converted to the electron energy to explain the GRB spectra by the
synchrotron radiation of electrons, but no promising mechanism of the
energy transfer has been proposed so far.\cite{asano03} If protons emit
coherent radiation, its frequency is about $\sim 3 \times 10^{11}
r_4^{-1} \gamma_1^{3} \Gamma_2$ Hz, which is less than the synchrotron
self-absorption frequency\cite{fan05} and also than the plasma frequency
of cold electrons. Therefore the radiation energy would be absorbed by
electrons efficiently and electrons would be heated up to the Lorentz
factor of $\gamma_e \sim (m_p/m_e) \gamma \sim 10^4 \gamma_1$. Then the
standard GRB model may be realized after heated electrons are
accelerated. Note that the electron heating should be relevant to the
injection problem of the acceleration.

The heated electrons may emit prompt GRBs by the inverse Compton
radiation, not by the synchrotron radiation. Interestingly, if the
coherent radiation from electrons is upscattered, the typical energy is 
$\sim \gamma_e^2 \nu_{*,{\rm obs}} \sim$ MeV, comparable to the peak
energy of the GRBs. In this scenario it is not necessary for the
magnetic field to survive for a long time.\cite{silva03,kato05}

\section{Discussions}

Above calculations are just order-of-magnitude estimates.
To establish the possibility of coherent radiation in relativistic
collisionless shocks, three-dimensional kinetic simulations are
necessary. In principle particle-in-cell
simulations\cite{nishikawa03,nishikawa05,jaroschek05} automatically
include the coherent effect, though these are challenging. To resolve
the wave length $\sim r/\gamma^3$ with at least $\sim 10$ grids and
cover the curvature radius $\sim r$, we need about $\sim [1000
(\gamma/5)^3]^3$ grids.

We have considered only the prompt GRBs as a first step. The coherent
radiation may be also important for the GRB afterglows, in which the
energy transfer from protons to electrons may depend on the Lorentz
factor $\gamma$ as implied by equation (\ref{eq:lNB}). Applications to
other astrophysical phenomena, such as AGN jets, microquasars, giant
flares from soft gamma-ray repeaters,\cite{ioka05} and pulsar wind
nebulae are also interesting.

Our simple model of a current filament does not include several
effects. We have neglected the velocity distribution parallel and
perpendicular to the propagation. The filament radius may be
varying. Other bunching mechanisms such as the sausage instability
may be more important than the radiative instability.
Also the Razin effect may be important since the frequency of
coherent radiation is relatively low. When we calculate the upscattered
coherent radiation, we may have to take the anisotropy of the radiation
field into account.\cite{ioka03} These are interesting future problems.

\section*{Acknowledgements}
We thank T.~N.~Kato, P.~M${\acute {\rm e}}$sz${\acute {\rm a}}$ros,
T.~Nakamura, K.~Toma and R.~Yamazaki for useful comments.
This work was supported in part by Grant-in-Aid of Monbukagaku-Sho
14047212 and 14204024.

%

\end{document}